# Chromospheric Sunspots in the Millimeter Range as Observed by the Nobeyama Radioheliograph

Kazumasa Iwai[1,2], Hideki Koshiishi[3], Kiyoto Shibasaki[2], Satoshi Nozawa[4], Shun Miyawaki[4], Takuro Yoneya[4]

1. National Institute of Information and Communications Technology, Koganei 184-8795, Tokyo, Japan; kazumasa.iwai@nict.go.jp
2. Nobeyama Solar Radio Observatory, National Astronomical Observatory of Japan, Minamimaki, Nagano 384-1305, Japan
3. Aerospace Research and Development Directorate, Japan Aerospace Exploration Agency, Tsukuba 305-8505, Japan
4. Department of Science, Ibaraki University, Mito, Ibaraki 310-8512, Japan

**ABSTRACT**

We investigate the upper chromosphere and the transition region of the sunspot umbra using the radio brightness temperature at 34 GHz (corresponding to 8.8-mm observations) as observed by the Nobeyama Radioheliograph (NoRH). Radio free-free emission in the longer millimeter range is generated around the transition region, and its brightness temperature yields the region's temperature and density distribution. We use the NoRH data at 34 GHz by applying the Steer-CLEAN image synthesis. These data and the analysis method enable us to investigate the chromospheric structures in the longer millimeter range with high spatial resolution and sufficient visibilities. We also perform simultaneous observations of one sunspot using the NoRH and the Nobeyama 45-m telescope operating at 115 GHz. We determine that 115-GHz emission mainly originates from the lower chromosphere while 34-GHz emission mainly originates from the upper chromosphere and transition region. These observational results are consistent with the radio emission characteristics estimated from the current atmospheric models of the chromosphere. On the other hand, the observed brightness temperature of the umbral region is almost the same as that of the quiet region. This result is inconsistent with the current sunspot models, which predict a considerably higher brightness temperature of the sunspot umbra at 34 GHz. This inconsistency suggests that the temperature of the region at which the 34 GHz radio emission becomes optically thick should be lower than that predicted by the models.

## 1. INTRODUCTION

The chromosphere, which is the second of the three main layers in the solar atmosphere, is known to be a structurally complex layer. Several studies have focused on the observation and modeling of the chromospheric atmosphere, and there are many empirical atmospheric models, including the one proposed by Vernazza et al. (1981). Other researchers have proposed several atmospheric models specific to sunspots (e.g., Avrett 1981; Maltby et al. 1986; Severino et al. 1994; Socas-Navarro 2007; Fontenla et al. 2009). These models are based on ultraviolet (UV), extreme ultraviolet (EUV), and optical observations of the chromosphere. In the chromosphere, the line broadening effect becomes significant because of the presence of hot and turbulent plasma. Hence, the acquisition of emission-line observations becomes more difficult than in the case of the photosphere. In addition, these emissions are formed under non-local thermodynamic equilibrium (non-LTE) conditions. Hence, the derivation of atmospheric models from the observational results requires complex radiative magnetohydrodynamic simulations.

The thermal bremsstrahlung, or so-called free-free emission, is emitted from the chromosphere in the submillimeter and millimeter ranges. These emissions are formed under the LTE condition. The opacity of the free-free emission depends on the emission measure and temperature (Dulk 1985). In addition, the Rayleigh–Jeans law is applicable for this wavelength range. Hence, we can estimate the radio brightness temperature from the vertical models of the atmospheric density and temperature. Certain previous studies have focused on sunspot observations in the submillimeter range (Bastian et al. 1993; Lindsey & Kopp. 1995) and the 3-mm range (White et al. 2006; Iwai & Shimojo 2015). These observations suggest that the brightness temperature of the sunspot umbra is lower than that of the quiet region when measured at wavelengths between the submillimeter and millimeter ranges. The current chromospheric models agree that the brightness temperature of the sunspot umbra should be lower than that of the quiet region in the submillimeter range, but predict that it should be higher than that of the quiet region in the 3-mm range (Loukitcheva et al. 2014). Hence, the observations and models are consistent over the submillimeter range and inconsistent for the 3-mm range. This inconsistency suggests that there may be some discrepancies of the models regarding the upper side of the chromosphere.

The free-free emission from the Sun at a given wavelength has a height distribution of the emission formation region (hereafter, we refer to this distribution as the emission contribution function). Hence, multi-frequency observations become important. In the longer millimeter range (5-10 mm), radio emission originates mainly from the upper chromosphere and the transition region (Wedemeyer-Böhm et al. 2007). In particular, the brightness temperature in the longer millimeter range is sensitive to the location of the transition region (Loukitcheva et al. 2014). Hence, the brightness temperature of this wavelength range is very important to evaluate the atmospheric structure.

There have been many solar observations in the longer millimeter band, such as those using the Hiraiso 10-m antenna at 9.5 mm (Kumagai et al. 1981) and those using the Nobeyama 45-m radio telescope at 8.3 mm (Kosugi et al. 1986; Irimajiri et al. 1995). These single-dish observations did not possess sufficient spatial resolutions to resolve the sunspots because the typical size of a sunspot umbra is smaller than 1′. We require a dish size of at least 100 m in the 9-mm band (the full width at half maximum is about 20″). Hence, the technique of interferometry is required to resolve the sunspot structures. In this context, Loukitcheva et al. (2014) have used the Very Large Array (VLA) at 15 GHz (20 mm). However, the VLA observations do not have sufficiently short baselines, which makes it difficult to measure the background intensity scale. On the other hand, a solar sunspot is a broad radio source that contains small structures, which requires many visibilities for image synthesis.

A more detailed comparison between atmospheric models and observations can be achieved if we can acquire observations over more wavelengths. In particular, the lack of observations in the longer millimeter range is significant, because these wavelengths are sensitive to the temperature and density distribution around the transition region, which is one of the least understood regions of the solar atmosphere. Hence, it is desired to derive the brightness temperature of the sunspots in the longer millimetric range with sufficient spatial resolutions and visibilities. The Nobeyama Radioheliograph (NoRH: Nakajima et al. 1994) is a radio interferometer that has 84 antennas operating at 34 GHz (8.8 mm). The longest baseline of the NoRH is about 489 m, which corresponds to a best spatial resolution of about 5″ at 34 GHz. The fundamental baseline is about 1.5 m. Therefore, the NoRH is suitable for observing a broad radio source that contains small structures such as solar sunspots. The CLEAN algorithm is generally applied to obtain "clean" images from the "dirty" images. This algorithm is suitable to deconvolve compact point sources such as flares (Hanaoka et al. 1994). On the other hand, the Steer-CLEAN algorithm can be used for diffuse sources such as active regions and filaments (Koshiishi 2003). The maximum entropy method (MEM) is used for large structures such as coronal holes and polar regions (Nindos et al. 1999). The CLEAN or Steer-CLEAN algorithms may be suitable for sunspot structures because the spatial scale of these structures is almost the same or larger than that of the beam size of the NoRH. However, there is no Steer-CLEAN option for the 34-GHz data in the current NoRH data analysis package.

The purpose of this study is to derive the brightness temperature of sunspot umbrae at 34 GHz using the NoRH. In our study, we applied the Steer-CLEAN algorithm to the 34-GHz data of NoRH. Subsequently, we estimated the emission regions of the millimetric radio waves and evaluated them in the context of the pre-existing atmospheric models. The instrument and data set used in this study are described in Section 2. The data analysis is presented in Section 3 and discussed in Section 4. We summarize this study in Section 5.

## 2. OBSERVATION

The Nobeyama Radioheliograph (NoRH: Nakajima et al. 1994) is a radio interferometer dedicated to solar observations. It has 84 antennas, each with a diameter of 80 cm. The NoRH observes the full solar disc every 1 s at 17 and 34 GHz. The NoRH uses the redundancy of the visibilities to calibrate the phase and amplitude of the instruments. In this study, we synthesize radio images every 1 s, which interval is shorter than the time variation of the phase. Next, the images are averaged to reduce statistical noise (Iwai et al, 2014); we average 120 images to obtain each averaged image. The brightness temperatures of the averaged images are normalized by the sky and the quiet Sun levels. We generate a histogram of the pixel counts of each image. The most frequent count level is defined as the background sky level, which is assumed to be at 0 K (we neglect the cosmic microwave background radiation). The second most frequent count level is defined as the quiet Sun level. The brightness temperature of the quiet Sun at 34 GHz is assumed to be 9000 K following Selhorst et al. (2005).

The three main processes that produce microwave and millimeter solar radio emission are free-free emission, gyro-resonance emission, and gyro-synchrotron emission. Gyro-resonance emission is emitted from sunspots at lower harmonics (mainly second or third) of their local gyro-frequency (Shibasaki et al. 1994; Nindos et al. 2000; Vourlidas et al. 2006). The third harmonic of the gyro-frequency from the magnetic field strength of about 4000 G corresponds to the observational frequency of this study (34 GHz). It is very rare that the magnetic strengths of sunspots exceed 4000 G. Hence, the gyro-resonance emission is negligible at this wavelength. Gyro-synchrotron emission is emitted from non-thermal electrons during flares. As regards the study, we examined the flare list created by Watanabe et al. (2012) and chose the data that contained no C-class or larger flares as observed by the Geostationary Operational Environmental Satellite (GOES).

The shortest baseline of the NoRH (1.5 m) is too long to detect the solar disc (~0.5 degree) at 34 GHz. Hence, the 34-GHz images usually contain dummy structures due to the spatial aliasing effect. In this study, we selected the data in which the aliasing structures do not overlap the sunspot structures by avoiding sunspots located near the limb, where the contamination of the aliasing disc components is relatively large. We also used the data obtained at lower elevations because their effective baseline becomes shorter, and the resulting aliasing effect becomes smaller. However, this means that the spatial resolution of the synthesized image becomes lower than the best situation. Hence, we used relatively large and isolated sunspots. Figure 1(a) shows the full disc image of the Sun at 34 GHz as observed on 2014 January 8. The aliasing dummy structures of the disc are indicated by the red dashed lines. The bright region marked by the red arrows indicates the aliasing dummy of the active region around the west limb, which is indicated by the white arrow. The active region analyzed in this study corresponds to the red rectangle. From the image, we note that the active region of interest was not affected by the spatial aliasing structures.

## 3. DATA ANALYSIS AND RESULTS

The Steer-CLEAN algorithm is a CLEAN algorithm proposed by Steer et al. (1984). It was applied to the NoRH data analysis package by Koshiishi (2003) at 17 GHz. In this study, we applied this algorithm to the processing of 34-GHz NoRH data. In this algorithm, a group of radio sources called contour-trims is subtracted from the dirty image with a CLEAN loop gain, while the standard CLEAN algorithm subtracts an individual radio source from the dirty image. We used values of 0.8 for the contour-trim level and 0.2 for the CLEAN loop gain, which are the same as those at 17 GHz as per Koshiishi (2003). We defined the noise level of the data as the standard deviation of the sky region in the dirty image, which corresponds to the same region surrounded by the white rectangles in Figure 1. Subsequently, we set a value of 3σ for the CLEAN criterion. The solar images observed on 2014 January 8, after synthesis by the CLEAN and Steer-CLEAN algorithms are shown in Figure 1. The bright region indicated by the red rectangles is one of the sunspot regions analyzed in this study. These figures show that the Steer-CLEAN synthesis provides a smoother image, and this effect is more apparent in the quiet regions in Figures 1(a) and (b). We used the images derived from the Steer-CLEAN algorithm in the following data analysis.

Figure 2 shows a sunspot region observed on 2014 January 8 at 34 GHz by NoRH and the UV and EUV wavelength ranges at 1700 Å, 304 Å, and 171 Å, as observed by the Atmospheric Imaging Assembly (AIA; Lemen et al. 2012) on board the Solar Dynamics Observatory (SDO). The sunspot umbra is indicated by the red rectangle in panel (a). The green contours indicate the radio brightness temperature of 9000 K, which is the quiet Sun level. We note that the sunspot umbra is darker than the surrounding plage region at 34 GHz. The brightness temperature of the sunspot umbra is almost the same as that of the quiet region.

Figure 3 shows the radio and UV/EUV images observed on 2014 October 24. There is a low-radio-brightness-temperature region around the sunspot umbra in Figure 3(b). However, the center of the sunspot umbra at 1700 Å appears to be slightly northward of the low-radio-brightness-temperature region. The brightness temperature of the actual center of this umbra region is higher than that of the quiet region.

Figure 4 shows the radio and UV/EUV images observed on 2014 February 12. This sunspot region was also observed by the Nobeyama 45-m telescope in the 3-mm range (Iwai and Shimojo 2015). This sunspot region is relatively smaller than the two abovementioned sunspots in Figures 2 and 3. The radio brightness temperature of the sunspot umbra is almost the same as that of the quiet region.

Table 1 summarizes the brightness temperatures of the sunspot umbra and quiet regions surrounded by the red and white rectangles, respectively, in Figures 2, 3, and 4. The brightness temperature of the 115-GHz image is also defined by the histogram of the observed level. The quiet region level at 115 GHz was set as 7500 K from Selhorst et al. (2005). Hence, the quiet region levels of the radio emissions are defined by the data of the entire disc. Therefore, the brightness temperatures of the quiet regions surrounded by the white rectangles are different from the quiet region values (9000 K at 34 GHz and 7500 K at 115 GHz).

In this study, we selected an additional 8 active regions observed by the NoRH at 34 GHz, in order to confirm the generality of this finding of low umbral contrast. The selected active regions contained relatively large and well-isolated sunspots including the sunspots observed on 2014 January 8 (see Figure 2) and 2014 October 24 (see Figure 3). We did not use the data from the sunspot observed on 2014 February 12 (Figure 4) because it was not large relative to the other sunspots. Although the NoRH observes the Sun every day, we limited the selection of regions to those with large sunspots located near the disc center for which SDO/AIA data are available. We also eliminated the data affected by flares and the spatial aliasing effect. We chose a 20″ square region at the center of the umbra for each sunspot and derived the averaged brightness temperature of the center regions. Table 2 lists the 10 selected sunspots.

## 4. DISCUSSION
### 4.1. Image quality

In this study, we applied two types of image synthesis, i.e., the CLEAN and Steer-CLEAN syntheses. Our result shows that the Steer-CLEAN synthesis provides a smoother image. This effect is more apparent in the quiet regions in Figures 1(a) and (b). The standard deviations of the brightness temperatures of the sky region surrounded by the white rectangles in Figures 1(a) and (b) are 377 K and 284 K, respectively. Hence, the Steer-CLEAN synthesis provides a slightly better image from the same NoRH data. However, the Steer-CLEAN synthesis provides smoother structures in the peak brightness region. Hence, the dynamic ranges of these images (the ratio between the peak brightness and the standard deviation of the sky region) are almost the same (~21 dB). It should be noted that the sky region surrounded by the white rectangles in Figure 1 partially contains the aliasing structures of the disc. Hence, the noise level of the sky region strongly varies with the region of our focus.

Although the best spatial resolution of the NoRH at 34 GHz is about 5″, the typical spatial resolutions in this study were larger (but smaller than 10″). This is because we chose relatively low elevation data to minimize the aliasing of the disc structures (see Figure 1). The umbra sizes in this study are close to or larger than 20″. Hence, the umbrae are sufficiently large to be resolved. Moreover, we derived similar images from the CLEAN and Steer-CLEAN syntheses. Therefore, we conclude that the derived structures in the radio images are real structures, and we use the Steer-CLEAN images for the following discussions.

**4.2 Region corresponding to 34-GHz and UV/EUV emissions**
In this study, we compared the radio images with three UV/EUV images. The image corresponding to emission at 1700 Å is generated around the temperature minimum and the photosphere, and its characteristic temperature (log T) is 3.7. The 304 Å emission is mainly generated in the chromosphere and transition region, and its characteristic temperature is 4.7. The 171 Å emission is generated in the quiet corona and upper transition region, and its characteristic temperature is 5.8 (Lemen et al. 2012).

Figures 2 and 3 indicate that there is a good agreement of the emission region between the 34-GHz radio image and the chromospheric emission at 304 Å over 171 Å, which suggests that 34-GHz emission mainly arises from the chromosphere. This is consistent with previous coronal radiative transfer simulations of the radio free-free emission. For example, Grebinskij et al. (2000) calculated the amount of the radio free-free emission at 34 GHz using a typical coronal atmospheric model. Their results suggest that the coronal component is less than 10% of the total radio emission at 34 GHz.

Figure 4 indicates that the emission region at 115 GHz matches that of 1700 Å, while the emission region at 34 GHz matches that of 304 Å. This result is also consistent with previous chromospheric radiative transfer simulations of the radio free-free emission. For example, Wedemeyer-Böhm et al. (2007) calculated the contribution function between 0.3 and 9 mm. Their results suggest that the emission around the 3-mm range originates from the lower chromosphere. Moreover, Loukitcheva et al. (2004) suggested that the emission at $\lambda \geq 8$ mm originates from the transition region. The emission region of 34 GHz shows more diffuse structures than that of 115 GHz. The spatial resolution of the NoRH at 34 GHz (~10″) is higher than that of the 45-m telescope at 115 GHz (~15″). Hence, the diffuse structures at 34 GHz are attributed to the difference of the spatial structures of the emission region. In fact, the emission region of 304 Å shows more diffuse structures than that of 1700 Å.

In the sunspot region, the radio emission at 34 GHz exhibits various features. In Figure 2, the brightness temperature of the umbra region is lower than the surrounding plage region and almost the same as that of the quiet region. On the other hand, the umbra region is as bright as the surrounding plage region in Figure 3. In Figure 4, the umbra region is not bright, and there are no characteristic structures that indicate the presence of an umbra. This difference in the radio emission characteristics at 34 GHz appears to be related to the loop structures of the 304 Å image. For example, the low-radio-brightness-temperature region around the sunspot in Figure 3 appears to correspond to the low-emission region for 304 Å in Figure 3(c). Figures 2-4 suggest that even if there is a large sunspot region, the umbra region appears bright if there are loop structures above the umbra region that generate EUV emission at 304 Å. This result indicates that the radio emission at 34 GHz includes not only upper chromomeric emission but also emission from the transition region and lower corona where the atmosphere is not stratified and various magnetic loop structures exist.

If the loop structures above the chromosphere emit 34-GHz emission, we need to consider the projection effect along the line of sight. In Figure 3, the active region is located in the southern hemisphere. The loop structures from the northern plage region are superposed on the umbra region in the line of sight. Hence, the observed brightness temperature in the umbra region may not be the actual brightness temperature above the umbra region.

**4.3 Statistical analysis**

Figure 5 shows the relationship between the radio brightness temperature and UV/EUV emissions of the sunspot umbrae listed in Table 2. In Figure 5(a), the radio brightness temperature of sunspot umbra is proportional to the EUV emission at 304 Å. This correlation suggests that the radio emission at 34 GHz contains emission from the loops above the sunspots if there are many overlying loops. On the other hand, the correlation between the radio and 304 Å emissions becomes weaker in sunspots with weak 304 Å emission. Hence, the effect of the overlying loops should be smaller in these sunspots. In Figure 5(b), the radio emission is also proportional to the 1700 Å emission except for one sunspot. This sunspot corresponded to the sunspot that contained the largest amount of 304 Å emission in Figure 5(a). The variation in 1700 Å emission is mainly due to the emission from the penumbra around the umbra and light bridges inside the umbra, which is bright at 1700 Å. Although we selected relatively large and isolated sunspots, we derived an average brightness within 20″ square regions that were larger than the beam size of the NoRH. Hence, it is difficult to fully eliminate the penumbra and light bridge regions. The relationship between 34-GHz emission and 1700 Å emission may suggest that the penumbra and light bridge regions might be bright in the radio wavelength range. However, detailed investigations must be carried out with the use of other instruments with higher spatial resolutions. At least some sunspots in Table 2 may contain emission from the surrounding penumbra and other fine structures included in the sunspots. From these discussions, it is suggested that the most accurate estimate of the sunspot brightness temperature can be derived from the sunspot that has the least amount of 1700 Å and 304 Å emissions, which is about 9000 K (same as the quiet Sun level). There should be a small amount of loop emission along with a minor gyro-resonance-emission component; however, these components are difficult to estimate. Therefore, we consider the observed brightness temperature of the umbra region as this region's upper limit.

The current atmospheric models of sunspots do not consider radio emission from the loop structures. A new model that includes the higher atmospheric layer and its magnetic structures will be required for future studies.

**4.4 Comparison with atmospheric models**
From the previous section, we recall that 34-GHz emission in the umbrae region is affected by the loop emission above the sunspots and emission from the penumbra and other fine structures of the sunspot. However, our statistical analysis suggests that the sunspot umbrae that contain the least amount of 304 and 1700 Å emissions can yield the upper limit of the brightness temperature of a sunspot umbra, which can be utilized as comparison data with respect to current atmospheric models.

We found that the upper limit of the brightness temperature of the umbra region was almost the same as that of the quiet region at 34 GHz. The previous millimetric observations suggested a similar conclusion in the 3-mm range (White et al. 2006; Iwai and Shimojo 2015). In the submillimeter-wavelength range, the brightness temperature of a sunspot umbra is clearly lower than that of the quiet region (Bastian et al. 1993; Lindsey & Kopp 1995). Loukitcheva et al. (2014) have calculated the brightness temperature of sunspot umbrae from various atmospheric models of sunspots (e.g., Avrett 1981; Maltby et al. 1986; Severino et al. 1994; Socas-Navarro 2007; Fontenla et al. 2009). Figure 6 reproduces Figure 8 in the study by Loukitcheva et al. (2014). Their findings suggest that sunspot umbrae should be darker than the quiet region in the submillimeter range, they should be brighter than the quiet region in the 3-mm range, and they should be considerably brighter than the quiet region in the 9-mm range. The red plus symbol in Figure 6 indicates our observational result. We also present the observational results by Iwai & Shimojo (2015) in the 3-mm range (black pluses). It is to be noted that the observed sunspot brightness temperatures form an upper limit, and the true brightness temperature could be even lower. From these observational and modeling results, the observed brightness temperature of the umbral region appears to be lower than that predicted by the atmospheric models in the millimeter range. This difference becomes larger in the longer millimeter range. Among the six atmospheric models represented in Figure 6, the model of Severino (1994) provides the most acceptable fit although it still predicts a brightness temperature that is greater than our observational upper limit.

Our findings indicate that 34-GHz emission occurs between the upper chromosphere and lower corona. This result is consistent with the atmospheric modeling results that suggest that the center of the contribution function of 34-GHz emission should lie around the upper chromosphere and the transition region. Hence, the difference between the radio observation and modeling of the sunspot umbra in this frequency range suggests an incomplete understanding of the transition region in the current sunspot atmospheric models. The lower brightness temperature suggests a lower temperature at the layer that the 34-GHz emission becomes "optically thick." This condition can be explained in many ways because the radio brightness temperature is a function of temperature and density, and the radio contribution function around 34 GHz is very broad. The atmospheric model by Severino (1994), which yields the most acceptable fit for our observations, predicts a lower brightness temperature in the 9-mm range as per the simulation by Loukitcheva et al. (2014), as shown in Figure 6 in this study. On the other hand, the atmospheric model by Fontenla (2009) predicts a higher brightness temperature at this wavelength range. The difference between these models in the transition region should generate the difference of the predicted radio brightness temperatures at 34 GHz. Loukitcheva et al. (2014) summarized the height distributions of the temperature and density of the atmospheric models (see Figure 6 in Loukitcheva et al. (2014)). In all models, the temperature increases steeply and density decreases steeply in the transition region. However, in the model by Severino (1994), the steep changes in the temperature and density start from a lower height than those in the model by Fontenla (2009). If the steep changes in the temperature and density start from a lower height than that of the model by Severino (1994), we may be able to derive a model that yields a more acceptable fit for our observations. We should mention here that there might be many other explanations for our results. For example, the model by Severino (1994) also sets an extended temperature minimum region relative to other models. This difference can also affect the emission in the long millimetric range because of the broad contribution function in this wavelength range. At least, our observational result indicates that the atmospheric heating above the sunspot umbra should be different from that expected as per current model predictions.

Lower-frequency observations such as the 17-GHz data of the NoRH or VLA can be used to acquire a clear picture of the atmospheric structure of the upper layer. For example, Loukitcheva et al. (2014) investigated the brightness temperature of two sunspots using VLA data at 15 GHz. They suggested that the expected high brightness temperature of the sunspots at lower frequencies was not observed. Hence, their results are consistent with our results. However, lower-frequency observations are subject to a relatively larger amount of loop emission, as suggested by Iwai et al. (2014). The gyro-resonance emission is also contained in the lower frequency. Hence, even if we can investigate more sunspots over longer wavelength ranges, it is difficult to determine the frequency at which the sunspot umbra becomes brighter than the quiet region. This is the limitation of the radio diagnostics of the solar atmosphere.

In this study, we determined that the characteristics of the brightness temperature of the sunspot umbra can differ from one sunspot to another due to the magnetic loop structures above the sunspot region. Because millimetric radio emission has a broad contribution function, this effect should extensively reflect in the millimetric range. Previous studies on this topic have utilized only limited observation frequencies. Even though there are some studies that are based on simultaneous multi-frequency observations (e.g., Iwai and Shimojo 2015), the observation frequencies in these cases were too close, and the emission regions of the observed frequencies were almost the same. Figure 4 in this study depicts the first result that indicates the presence of millimetric radio emission generated simultaneously from two different layers. However, we have only one sunspot data set that contains both 115- and 34-GHz observations. Hence, we cannot address the brightness temperature spectra among the different sunspots. A simultaneous spectral observation will be important for future studies in this regard.

## 5. SUMMARY

This study investigated the umbral regions of sunspots using the Nobeyama Radioheliograph operating at 34 GHz. We applied the Steer-CLEAN image synthesis for the 34-GHz data. We found that the deep CLEAN using the Steer algorithm yields broad structures of non-flaring regions. The NoRH data and the Steer-CLEAN analysis method enable us to investigate the chromospheric structures in the longer millimeter range (8.8 mm) with high spatial resolution and sufficient visibilities for the first time.

We found that the emission region at 34 GHz matches that of 304 Å. In addition, we showed that 115-GHz emission mainly originates from the lower chromosphere and that 34-GHz emission mainly originates from the upper chromosphere and transition region using simultaneous observations made between NoRH and the Nobeyama 45-m telescope. These observational results are consistent with the radio emission characteristics estimated from the current atmospheric models of the chromosphere.

Our observational result indicates that the sunspot regions are affected by the loop structures of the surrounding plage regions. On the other hand, our statistical analysis suggests that an isolated sunspot region can provide the upper limit of the brightness temperature of the umbra region. The observed upper limit of the brightness temperature of the umbra region was nearly identical to that of the quiet region. This result is inconsistent with the current sunspot models that predict a considerably higher brightness temperature of the sunspot umbra at 34 GHz. This inconsistency suggests that the temperature of the region at which the 34-GHz radio emission becomes optically thick should be lower than that predicted by the models.

This study limits the observational targets to certain large sunspot regions because of the limitations of the spatial resolutions of the instruments utilized in the study. However, a large sunspot usually has a high activity, and thus, it is difficult to investigate the non-flare characteristics such as the sunspot umbra regardless of the spatial resolutions of the instruments. Moreover, this study suggests that the brightness temperature of the sunspot umbra can differ from one sunspot to another. For further investigation, simultaneous observation at multi-frequency bands with higher spatial resolutions will be required, which can be achieved by the Atacama Large Millimeter/submillimeter Array (ALMA).

ACKNOWLEDGEMENTS

This study is based on the results obtained from the Coordinated Data Analysis Workshop, 2014 (CDAW2014), organized by the Nobeyama Solar Radio Observatory (NSRO), the National Astronomical Observatory of Japan (NAOJ), and the Solar-Terrestrial Environment Laboratory (STEL), Nagoya University. The NSRO was shut down in 2015, and the Nobeyama Radioheliograph is now operated by the International Consortium for the Continued Operation of Nobeyama Radioheliograph (ICCON). ICCON consists of STEL/Nagoya University, NAOC, KASI, NICT, and GSFC/NASA. NoRH will be operated by these organizations. Then, we will realize that the quality of the observation is not determined by the affiliations of the operators but by their passion for better science. We thank NASA/SDO and the AIA and HMI science teams for the SDO data. The Nobeyama 45-m radio telescope is operated by Nobeyama Radio Observatory, NAOJ.

**Figures and Tables**

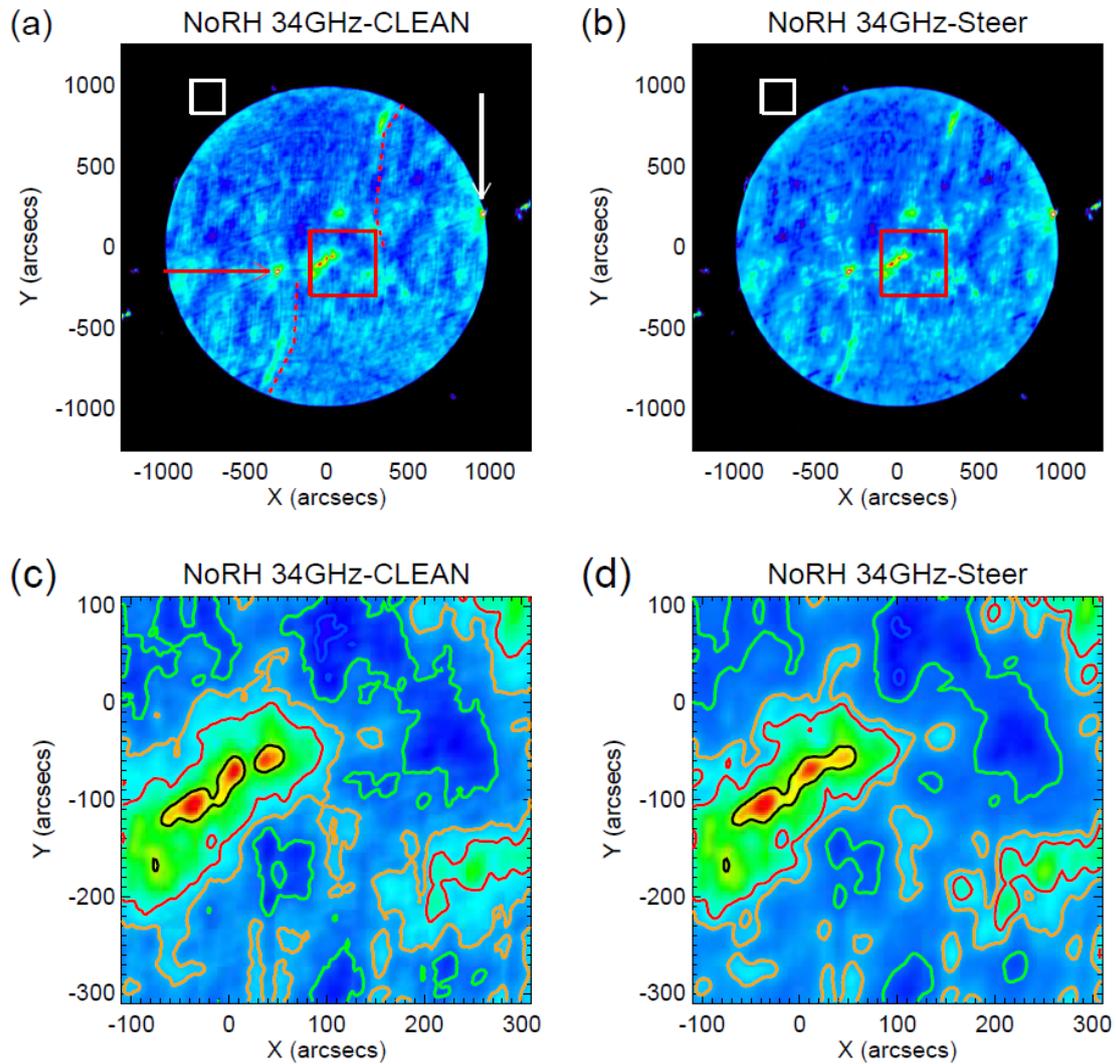

**Figure 1** Solar disc images observed by Nobeyama Radioheliograph (NoRH) at 34 GHz on 2014 January 8, synthesized by (a) the CLEAN and (b) Steer-CLEAN algorithms. The active region indicated by the red rectangles in panels (a) and (b) is shown in panels (c) and (d), respectively. The overlaid color contours include black: 18000 K, red: 11000 K, orange: 10000 K, green: 9000 K, and blue: 8000 K.

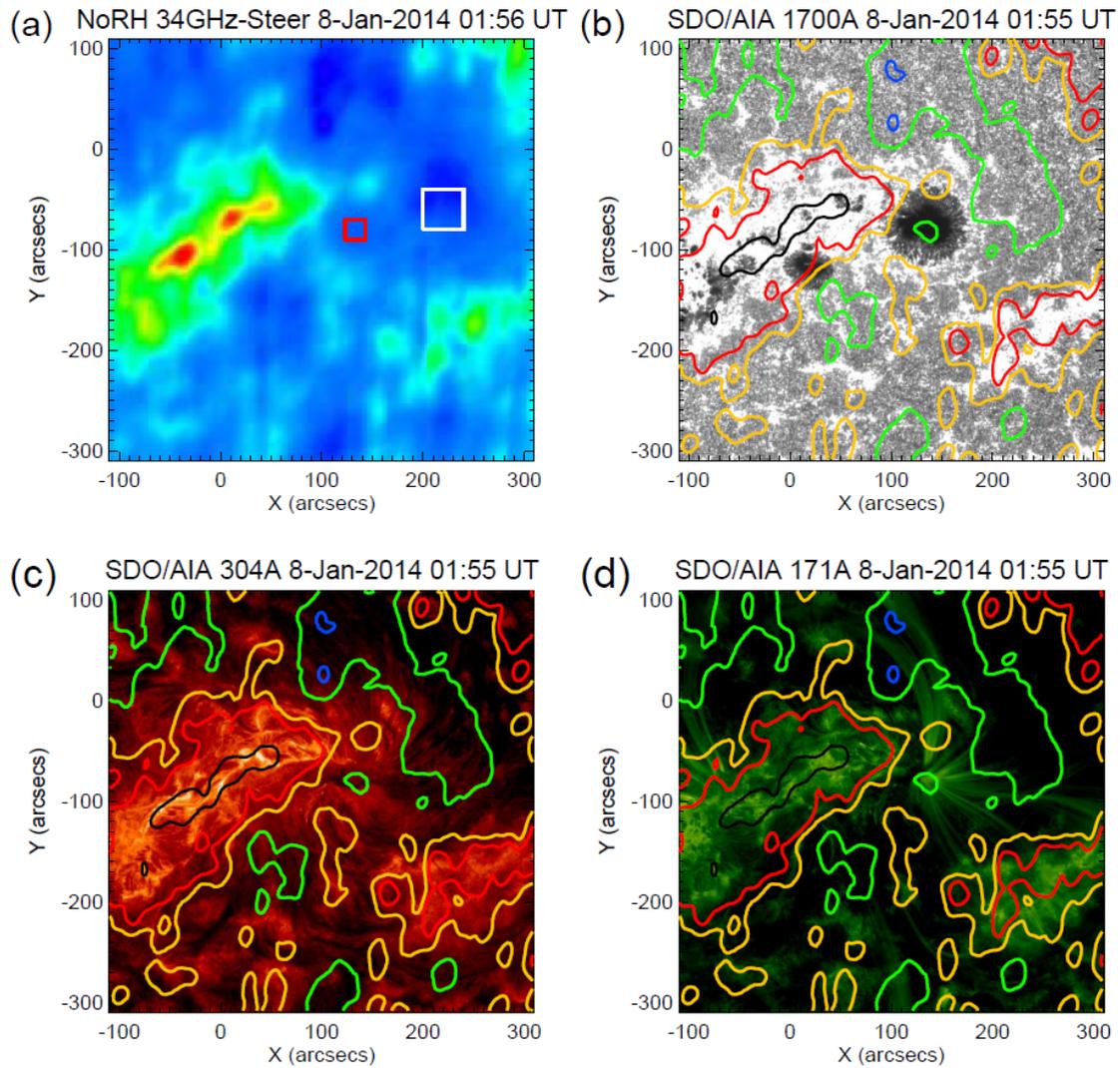

**Figure 2** (a) Radio image of the sunspot region observed by NoRH at 34 GHz using the Steer-CLEAN algorithm on 2014 January 8. Radio contour map observed at 34 GHz overlaid on (b) the 1700 Å, (c) 304 Å, and (d) 171 Å images acquired by the AIA. Black: 18000 K, red: 11000 K, orange: 10000 K, green: 9000 K, and blue: 8000 K.

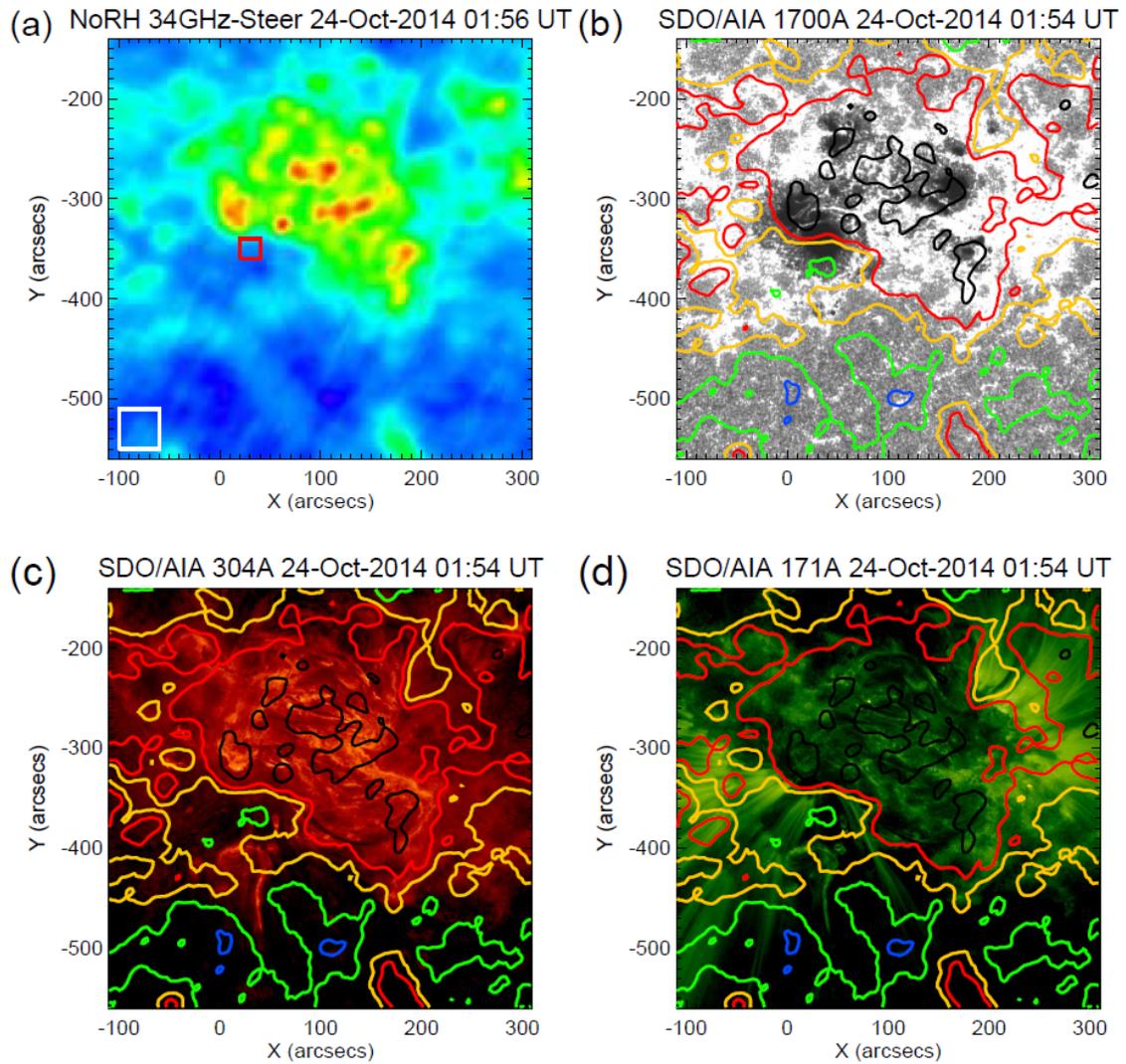

**Figure 3** (a) Radio image of the sunspot region observed by NoRH at 34 GHz using the Steer-CLEAN algorithm on 2014 October 24. Radio contour map observed at 34 GHz overlaid on (b) the 1700 Å, (c) 304 Å, and (d) 171 Å images acquired by the AIA. Black: 18000 K, red: 11000 K, orange: 10000 K, green: 9000 K, and blue: 8000 K.

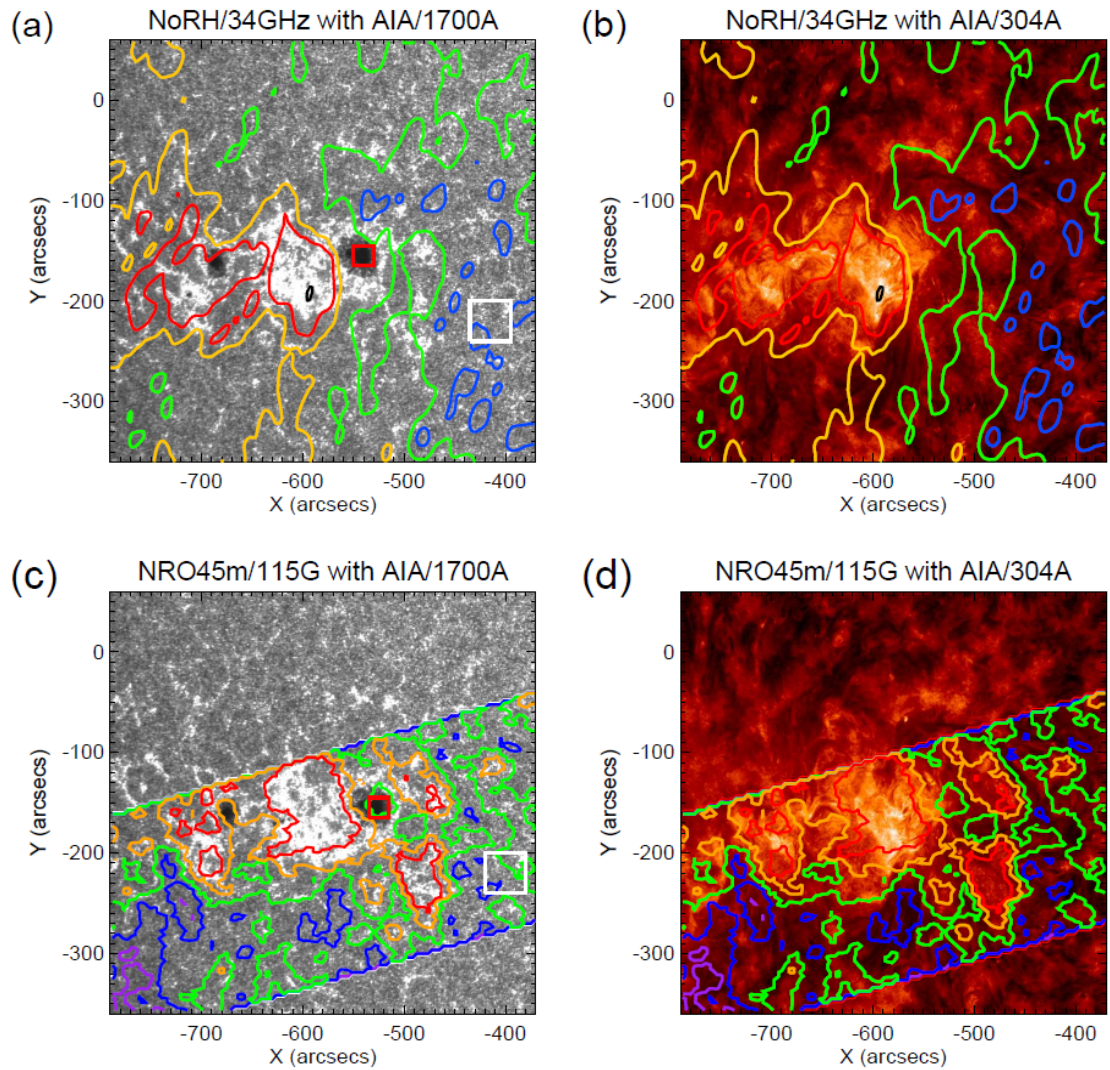

**Figure 4** Radio contour map image of the sunspot region observed by NoRH at 34 GHz using the Steer-CLEAN algorithm on 2014 February 12 overlaid on the (a) 1700 Å and (b) 304 Å images acquired by the AIA. Black: 18000 K, red: 11000 K, orange: 10000 K, green: 9000 K, and blue: 8000 K. Radio contour map observed at 115 GHz by the Nobeyama 45-m radio telescope overlaid on the (c) 1700 Å and (d) 304 Å images. Red: 7900 K, orange: 7700 K, green: 7500 K, blue: 7300 K, and purple: 7100 K.

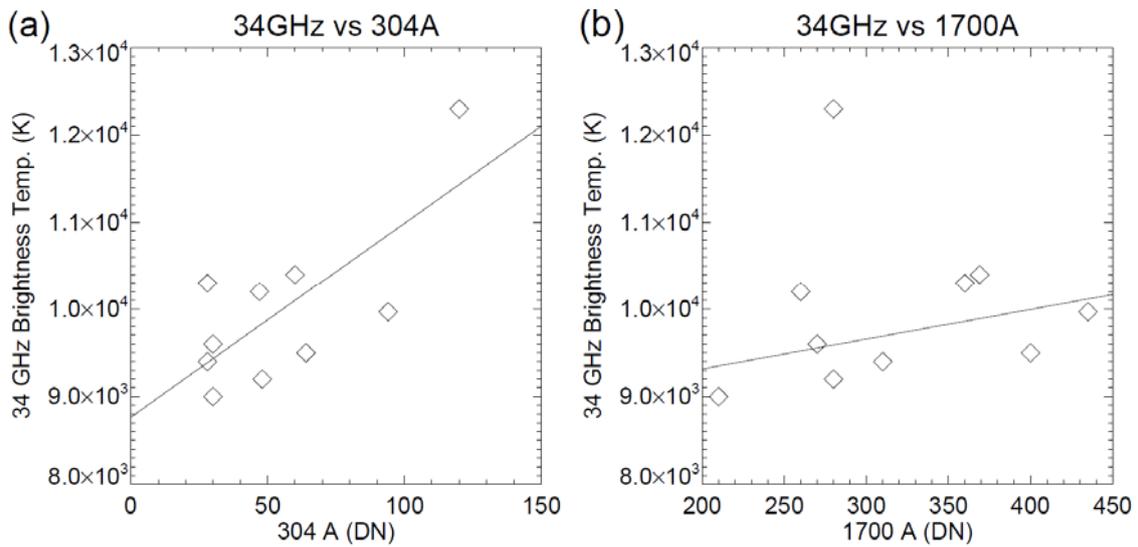

**Figure 5** Relationship between radio emission at 34 GHz and (a) 304 Å and (b) 1700 Å on the sunspot umbra listed in Table 2. Solid lines show the linear fitting results. The sunspot that contained a large amount of 304 Å emission was eliminated in the fitting in panel (b)

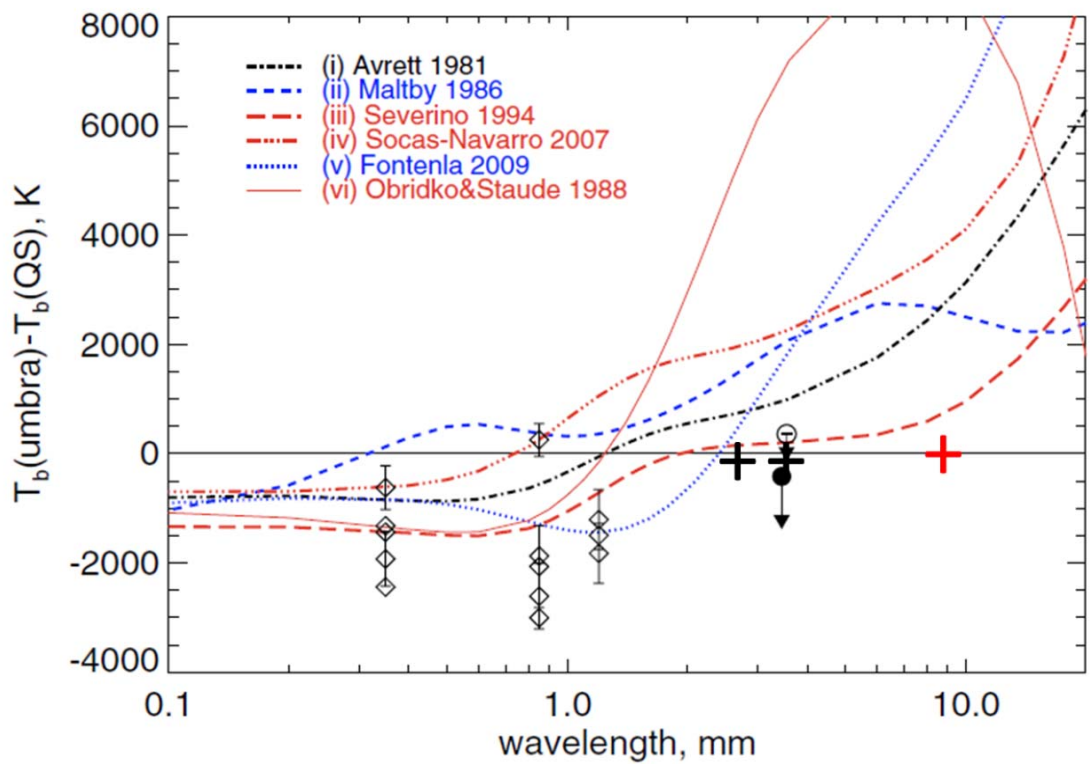

**Figure 6**

The umbral brightness temperature relative to the quiet Sun plotted as a function of wavelength based on Loukitcheva et al. (2014). The six lines indicate the predicted brightness from the models of (i) Avrett (1981), (ii) Maltby et al. (1986), (iii) Severino et al. (1994), (iv) Socas-Navarro (2007), (v) Fontenla et al. (2009), and (vi) Obridko & Staude (1988). The symbols indicate the observational values (filled and open circles) obtained from BIMA at 3.5 mm by White et al. (2006), (diamonds) from JCMT at 0.35, 0.85, and 1.2 mm obtained by Lindsey & Kopp (1995), (black plus) from the Nobeyama 45-m telescope at 2.6 and 3.5 mm obtained by Iwai & Shimojo (2015), and from NoRH at 8.8 mm (red plus, this study).

**Table 1** Radio brightness temperatures of the umbra and quiet regions surrounded by red and white rectangles, respectively, in Figures 2, 3, and 4.

| Day | Wavelength | Umbra | Quiet region |
|---|---|---|---|
| 20140108 | 34 GHz (K) | 9000 | 8600 |
| 20141024 | 34 GHz (K) | 9600 | 9100 |
| 20140212 | 34 GHz (K) | 9300 | 8300 |
|  | 115 GHz (K) | 7460 | 7440 |

**Table 2** Radio brightness temperatures and UV/EUV emissions at 1700 and 304 Å of the 10 umbra.

| Day | AR No. | 34 GHz (K) | 1700 (DN) | 304 (DN) |
|---|---|---|---|---|
| 2011/9/13 | AR11289 | 10000 | 430 | 90 |
| 2011/9/28 | AR11302 | 9500 | 400 | 60 |
| 2012/7/12 | AR11520 | 12300 | 280 | 120 |
| 2013/11/19 | AR11189 | 10400 | 370 | 60 |
| 2014/1/8 | AR11944 | 9000 | 210 | 30 |
| 2014/5/11 | AR12055 | 10200 | 260 | 50 |
| 2014/7/8 | AR12109 | 9200 | 280 | 50 |
| 2014/10/24 | AR12192 | 9600 | 270 | 30 |
| 2014/12/2 | AR12222 | 9400 | 310 | 30 |
| 2015/8/8 | AR12396 | 10300 | 360 | 30 |